\documentclass[12pt]{revtex4}
\begin{document}
\title{Finite nilpotent symmetry in Batalin-Vilkovisky formalism}


\author{Bhabani Prasad Mandal \footnote{e-mail address:
\ \ bhabani@bhu.ac.in, \ \ bhabani.mandal@gmail.com  }}\author{ Sumit Kumar Rai \footnote{e-mail address: sumitssc@gmail.com}}\author{Sudhaker Upadhyay \footnote{sudhakerupadhyay@gmail.com}}


\affiliation{ Department of Physics,\\
Banaras Hindu University,\\
Varanasi-221005, INDIA. \\
}

\begin{abstract}
We consider the Batalin-Vilkovisky  formulation of both 1-form and 2-form gauge theories, in the context of generalized BRST transformations with finite field dependent parameter. In the usual Faddeev-Popov formulation of gauge theories such finite field dependent BRST (FFBRST) transformations do not leave the generating functionals invariant as the path integral measure changes in a non-trivial way for a finite transformations. Here we show that FFBRST transformation, with appropriate choice of finite field-dependent parameter, is  symmetry of the generating functionals in the Batalin-Vilkovisky formalism. The finite parameter is chosen in such a way that the contribution from the Jacobian of the path integral measure is adjusted with gauge fixed fermions which do not change the generating functionals. Several examples for such a finite parameters are constructed.
\end{abstract}
\maketitle
The field/antifield formulation, alternatively known as  Batalin-Vilkovisky (BV) formalism \cite{bv1,bv2,bv3,gopa}, is one of the most powerful techniques to study the gauge field theories. This formulation is developed in Lagrangian framework and extremely useful as it allows us to deal with very general gauge theories including those with open or reducible gauge symmetry algebras.  The essential aspects of BV-formalism were originally developed by Zinn-Justin \cite{zz} in order to prove the renormalizability of gauge theories. This method provides a convenient way of analyzing the possible violations of symmetries of the action by quantum effects. This formulation
is based on the  BRST symmetry \cite{brs} which plays a crucial role in the discussion of quantization, renormalization, unitarity and other aspects of gauge theory. The nilpotent BRST transformation is characterized by an infinitesimal, anticommuting and space time independent parameter which leaves the FP effective action as well as the path integral measure in the generating functional invariant. 

In this present work, we show that FFBRST transformations \cite{jm} with appropriate choices of finite parameter are symmetry of the generating functional in the BV formulation.  In usual FP formulation, FFBRST transformations do not leave the generating functionals invariant as the path integral measure in the expression of generating functionals changes non-trivially under finite transformations. We choose the finite parameter in such a way that contribution from Jacobian for the path integral measure is adjusted with gauge fixed fermion which do not affect the generating functional in BV formulation. We construct few  finite parameter and show the results in the context of both 1-form and 2-form gauge theories. Generalized anti-BRST transformations are also constructed and are shown to be the symmetry of the generating functional in BV formulation of 1-form and 2-form gauge theories with appropriate choice of finite parameter. In principle, one can construct infinitely many such finite parameter for which FFBRST transformations  are symmetry of the generating functionals in BV formulation. 

The
main idea of BV formalism is to construct an extended action $W_\Psi(\Phi , \Phi^\star)$ by 
introducing antifields $\Phi^\star$ corresponding to each  field $\Phi$
with opposite statistic.  The sum of ghost number associated to a field and its antifield 
is equal to -1. Generically $\Phi $  denotes all the 
fields involved in the theory.  The generating functional can be written as 
\begin{equation}
 Z[\Phi^\star]= \int D{\Phi}e^{i W_{\Psi}\left[\Phi,\Phi ^*={\frac {\partial \Psi}{\partial {\Phi}}}\right].},
\end{equation}
$\Psi$ is the gauge fixed fermion and has Grassman parity 1 and ghost number {-1}. The
 generating functional $Z[\Phi^\star]$ is proved to be independent of the choice of  $\Psi$ \cite{ht}.
This extended quantum action satisfies certain rich mathematical
 relation called quantum master equation \cite{wei} and is  given by
\begin{equation}
\Delta e^{iW_\Psi[\Phi, \Phi^*]} =0  \ \mbox{ with }\ 
 \Delta\equiv \frac{\partial_r}{
\partial\Phi}\frac{\partial_r}{\partial\Phi^*} (-1)^{\epsilon
+1}.
\label{mq}
\end{equation}
Master equation reflects the gauge symmetry in the zeroth order of antifields
and in the  first order of antifields it reflects nilpotency of BRST transformation.
This equation  can also be written in terms of antibrackets as
\begin{equation}
\left ( W_\Psi, W_\Psi \right ) = 2i\Delta W_\Psi,
\label{ab}
\end{equation}
where the antibracket is defined as 
\begin{equation}
\left (X,Y\right ) \equiv \frac{\partial _rX}{\partial\Phi}\frac{\partial_l
Y}{\partial\Phi^*}-\frac{\partial _rX}{\partial\Phi^*}\frac{\partial_l
Y}{\partial\Phi}.
\label{ab1}
\end{equation}

Invariance of FP effective action does not depend on whether the BRST parameter is finite or field dependent as long as it is anticommuting and space-time independent. Keeping this in mind, Joglekar and Mandal have generalized the BRST symmetry by considering the anticommuting BRST parameter finite and field dependent \cite{jm}. Such a finite transformations relates the generating functionals corresponding to different effective theories in FP formulation \cite{rama}. The path integral measure is not invariant under such transformations as the parameter is finite and field dependent. It has been shown that Jacobian can be exponentiated to modify the effective action. Because of this interesting properties, FFBRST has found many applications \cite{sdj0,cou,sdj,rb,sdj1,etc}.
FFBRST transformations are
obtained by  integrating the infinitesimal  (field dependent ) BRST
transformations  \cite{jm}. In this method all the fields are allowed to be functions
of some parameter, $ \kappa : 0\le \kappa \le 1$. For a generic field $ \phi (x, \kappa),\
\phi(x, \kappa =0 ) = \phi(x) $, is the initial field and
 $ \phi(x, \kappa=1) = \phi ^\prime
(x)$ is the transformed field. Then the infinitesimal field dependent BRST transformations are
defined as
\begin{equation}
\frac{ d}{d \kappa}\phi(x, \kappa ) = \delta _{BRST }\ \phi(x, \kappa )\
\Theta ^ \prime [\phi(x,\kappa )],
\label{ibr}
\end{equation}
where $\Theta ^\prime d \kappa $ is an infinitesimal field dependent parameter.
It has been shown \cite{jm} by integrating these equations from $ \kappa=0$ to $\kappa=1$
that $\phi^\prime ( x) $ are related to $\phi(x)
$ by FFBRST transformation
\begin{equation}
\phi^\prime (x) = \phi(x) + \delta _{BRST} \ \phi(x)\ \Theta [\phi(x)],
\label{fbrs}
\end{equation}
where $ \Theta [\phi(x)] $ is obtained from $\Theta ^\prime [\phi(x)] $
through the relation
\begin{equation}
\Theta [\phi(x)] = \int^1_0 \Theta ^\prime [\phi(x)] d\kappa.
\label{80}
\end{equation}
Following  exactly the similar methods one can also construct finite field dependent anti-BRST (FFanti-BRST) transformations as
\begin{equation}
\phi^\prime (x) = \phi(x) + \delta _{anti-BRST} \ \phi(x)\ \Theta [\phi(x)].\label{fanti}
\end{equation}
These transformations are nilpotent and symmetry of the FP effective action. 
Now we are at a position to see the role of such transformations in BV formulation. In this present work, we cast the FFBRST transformations with appropriate choice of parameter as an formal nilpotent symmetry of the generating functional in the BV formulation. We show the results explicitly in BV formulation of both 1-form and 2-form gauge theories.\\
\\{\bf{\it In 1-form gauge theory: Yang-Mills theory:}}\\
\\We start with the generating functional of Yang-Mills theory in field/antifield formulation which can be written as
\begin{eqnarray}
Z[A_\mu^{\alpha\star},c^{\alpha\star},{\bar{c}}^{\alpha\star},B^{\alpha\star}] &=& \int [dAdcd\bar{c} dB] \exp\left [i\int d^4x \left \{-\frac{1}{4}F^{\alpha\mu\nu}F^\alpha_{\mu\nu}+A^{\mu\alpha*} D^{\alpha\beta}_\mu c^\beta \right.\right. \nonumber \\
&+&\left.\left. c^{\alpha *}\frac{g}{2}f^{\alpha\beta\gamma}c^\beta c^\gamma + B^\alpha \bar{c}^{\alpha *} \right\}\right],
\end{eqnarray}
or, compactly
\begin{equation}
Z[\Phi^*] = \int D\Phi \exp{\left [iW_\Psi(\Phi, \Phi^*)\right ]}\label{zla},
\end{equation}
where 
\begin{equation}
W_{\Psi}=S_0(\Phi)+\delta_{brst}\Psi.
\end{equation}
$\Psi $ is the gauge fixed fermion and can be written in this case as
\begin{equation}
\Psi =\int d^4 x \;\bar{c}^\alpha\left [\frac{\lambda}{2} B^\alpha - \partial\cdot A^\alpha \right ].
\end{equation}
The antifields $\Phi^*$ corresponding to the generic field
$\Phi$ are obtainable from the gauge fixed fermion as 
\begin{eqnarray}
\Phi^*=\frac{\delta\Psi}{\delta\Phi}.
\end{eqnarray}
Now we apply FFBRST transformation given in Eq. (\ref{fbrs}) with the finite field dependent parameter  $\Theta(A,c,\bar{c},B)$  
obtainable from 
\begin{equation}
\Theta ^\prime (A,c,\bar{c},B)= i\int d^4 y\;\bar{c}^\alpha \left[\gamma_1\lambda B^\alpha +(\partial\cdot A^\alpha -\eta
\cdot A^ \alpha)\right], \label{ta}
\end{equation}
using Eq. (\ref{80}) to the extended generating functional given in Eq. (\ref{zla}). Note even though the parameter, $\Theta$ is finite and field dependent, it is anticommuting in nature. The path integral measure in the expression of $Z[\Phi^*]$ is not invariant under such finite field dependent transformations and the generating functional change to 
\begin{equation}
Z[\tilde{\Phi}^*] = \int D\Phi \exp{\left [iW_{\Psi_1}(\Phi, \tilde{\Phi}^*)\right ]},
\end{equation}
where
\begin{equation}
W_{\Psi_1}=S_0(\Phi)+\delta_{brst}\Psi_1 ,
\end{equation}
with $\tilde{\Phi}^*=\frac{\delta\Psi_1}{\delta\Phi}$. The  gauge fixed fermion is changed from $\Psi\rightarrow\Psi_1$ as
\begin{equation}
\Psi_1 =\int d^4 x \;\bar{c}^\alpha\left [\frac{\xi}{2} B^\alpha - \eta\cdot A^\alpha \right ],
\end{equation}
with $\xi=\;\lambda(1+2\gamma)$.  However $Z[\tilde{\Phi}^*]$ is independent of the choice of $\Psi$ as proved in Ref. \cite{ht}. Thus, we see that nilpotent, finite BRST transformations
\begin{equation}
\phi^\prime (x) = \phi(x) + \delta _{BRST} \ \phi(x)\ \Theta [\phi(x)],
\end{equation}
with the finite parameter given in Eq. (\ref{ta}) along with $\tilde{\Phi}^*\rightarrow{\Phi}^*$ is a formal symmetry of generating functional $Z[\Phi^*]$ in BV formulation. We can construct many different choices of the finite field dependent parameter $\Theta$ which changes only the gauge fixed fermion through a non-trivial Jacobian in the path integral measure and hence are  the formal symmetry of the generating functional. For example, we make another choice of finite field dependent BRST parameter  as
\begin{equation}
\Theta^\prime (A,c,\bar{c},B)= i\int d^4 y \;\bar{c}^
\alpha \left[\gamma_1 \lambda B^\alpha + \partial_0 A^\alpha_0 \right],\label{tc}
\end{equation}
the generating functional $Z[\Phi^\star]$ changes to
\begin{equation}
Z[\tilde{\Phi}^*] = \int D\Phi \exp{\left [iW_{\Psi_2}(\Phi,\tilde{\Phi}^*)\right ]},
\end{equation}
where $\Psi_2$ is  the gauge fixed fermion given as 
\begin{equation}
\Psi_2 =\int d^4 x \;\bar{c}^\alpha\left [\frac{\xi}{2} B^\alpha - \partial^i A^\alpha_i \right ],
\end{equation}
\begin{equation}
Z[\Phi^*]\stackrel{\Theta ^\prime}{\stackrel{------------\longrightarrow}{FFBRST with}}Z[\tilde{\Phi}^*]
\end{equation}
Thus, the FFBRST transformations with the parameter given in Eq. (\ref{tc}) is also the formal nilpotent symmetry of the generating functional in BV formulation.\\
\\{\bf{\it In 2-form gauge theory:}}\\
\\
In this section, we consider BV formulation of Abelian 2-form gauge theories which is extremely useful in the study of string theories, dual formulation of Abelian Higgs model, supergravity theories with higher curvature term \cite{kara,suga,suga1,sase,grsc,grsc1,crsc,crsc1,crsc2,crsc3,crsc4,frne,frne1,frne2,dema,suma}. We show that the FFBRST transformations with a careful choice of finite parameter are the symmetry of the generating functional of 2-form gauge theory in BV formulation.
We start with effective action for Abelian gauge theory for rank-2 antisymmetric tensor field $B_{\mu\nu}$ 
defined as \cite{dema}
\begin{eqnarray}
S&=&\int d^4x \left[\frac{1}{12}F_{\mu \nu \rho}F^{\mu \nu \rho}-i\partial_\mu\tilde\rho_\nu (\partial^\mu\rho^\nu -
\partial^\nu\rho^\mu )+\partial_\mu\tilde\sigma\partial^\mu\sigma +\beta_\nu(\partial_\mu B^{
\mu\nu} +\lambda_1\beta^\nu -\partial^\nu\varphi)\right.\nonumber\\ 
&-&\left. i\tilde\chi\partial_\mu\rho^\mu -i\chi (\partial_\mu\tilde\rho^\mu -
\lambda_2\tilde\chi)\right], 
\end{eqnarray}
where $F_{\mu\nu\lambda}=\partial_\mu B_{\nu\lambda}+\partial_\nu B_{\lambda\mu}+\partial_\lambda B_{\mu\nu}$, $B_{\mu\nu}$ is the antisymmetric  tensor field of rank-2, ($\rho_\mu,\tilde{\rho}_\mu $) are anticommuting vector fields (ghost),
 ($\sigma_\mu,\tilde{\sigma}_\mu )$ are commuting scalar field,
 ($\chi,\tilde{\chi}$) are anticommuting scalar fields, and 
 ($\beta_\mu$, $\varphi$) are commuting vector 
 and scalar field respectively.
The generating functional  for this theory in BV formulation can be written as
\begin{eqnarray}
Z\left[B^{\mu\nu *},\rho^{\mu *},\tilde{\rho}^{\mu*},\tilde{\sigma}^*,\varphi^*\right] &=&\int\left[dBd\rho d\tilde{\rho}d\sigma d\tilde{\sigma}d\varphi d\chi d\tilde{\chi}d\beta\right]\exp\left[i\int d^4x\left\{\frac{1}{12}F_{\mu\nu\lambda}F^{\mu\nu\lambda}\right.\right.\nonumber\\
&-&B^{\mu\nu\star}\left(\partial_\mu\rho_\nu-\partial_\nu\rho_\mu\right)
-i\left.\left.\rho^{\mu\star}\partial_\mu\rho +i{\tilde{\rho}}^{\nu\star}\beta_\nu-\tilde{\sigma}^\star\tilde\chi-\varphi^\star\chi\right\}\right].
\end{eqnarray}
This can be expressed compactly as
\begin{equation}
Z[\Phi^*] = \int D\Phi \exp{\left [iW_{\Psi_3}(\Phi,\tilde{\Phi}^\star)\right]},\label{gf2f}
\end{equation}
where 
\begin{equation}
W_{\Psi_3}=S_0(\Phi)+\delta_{brst}\Psi_3.
\end{equation}
$\Psi_3 $ is  the gauge fixed fermion given as 
\begin{equation}
\Psi_3 =-i\int d^4x \left[\tilde\rho_\nu \left(\partial_\mu B^{\mu\nu}+
\lambda_1\beta^\nu \right) +\tilde\sigma \partial_\mu\rho^\mu +\varphi \left(\partial_\mu\tilde{\rho}^
\mu-\lambda_2\tilde\chi \right)\right].
\end{equation}
The antifields $\Phi^\star$ corresponding to generic field $\Phi$ for this particular theory can be obtained from the gauge fixed fermion using
$\tilde{\Phi}^\star=\frac{\delta\Psi_3}{\delta\Phi}$.
Now we choose a FFBRST parameter corresponding to
\begin{eqnarray}
\Theta^\prime &=&\int d^4x\left[\gamma_1\tilde\rho_\nu (\partial_\mu B^{\mu\nu}-\eta_\mu B^{
\mu\nu}-\partial^\nu\varphi -\eta^\nu\varphi )+\gamma_2 \lambda_1\tilde\rho_\nu\beta^\nu 
\right. \nonumber\\
&+&\left.\gamma_1\tilde\sigma (\partial_\mu\rho^\mu -\eta_\mu\rho^\mu )+\gamma_2 
\lambda_2\tilde\sigma\chi\right]\label{tp2f}
\end{eqnarray}
and apply FFBRST transformations to the generating functional given in Eq. (\ref{gf2f}). This takes Z$[\Phi^\star]$ to $Z[\tilde{\Phi}^\star]$ where
\begin{equation}
Z[\tilde{\Phi}^\star]=\int D\phi \exp[iW_{\Psi_4}(\Phi,\tilde{\Phi}^\star],
\end{equation}

\begin{equation}
W_{\Psi_4}=S_0(\Phi)+\delta_{brst}\Psi_4,
\end{equation}
and
\begin{equation}
\Psi_4 =-i\int d^4x \left[\tilde\rho_\nu \left(\eta_\mu B^{\mu\nu}+
\lambda_1\beta^\nu \right) +\tilde\sigma \eta_\mu\rho^\mu +\varphi \left(\eta_\mu\tilde{\rho}^
\mu-\lambda_2\tilde\chi \right)\right].
\end{equation}
with the corresponding antifields as $\tilde{\Phi}^\star=\frac{\delta{\Psi_4}}{\delta\Phi}$.
Thus, the FFBRST transformation in Eq. (\ref{fbrs}) with the parameter $\Theta^\prime$ given in Eq. (\ref{tp2f}) is a formal symmetry of the generating functional, $Z[\Phi^\star]$.
In fact, the FFBRST transformations corresponding to any choice of $\Theta^\prime$ which changes one gauge fixed fermion to another gauge fixed fermion through a non-trivial Jacobian of the path integral measure are also symmetry in the BV formulation of Abelian 2-form gauge theory.
\linebreak

{\it {Anti-BRST:}}\\
Anti-BRST transformations are analogous to BRST transformations where the role of ghosts and anti-ghosts fields are interchanged apart from some numerical factors.
Formal nilpotent symmetry transformations for the generating functionals in BV formulation can also be constructed using FFanti-BRST transformations 
[Eq. (\ref{fanti})]. For example, FFanti-BRST transformations with finite field parameter corresponding to
\begin{equation}
\Theta^\prime =-i\gamma \int d^4x\ c^\alpha (\partial\cdot A^\alpha -\partial_j A^{j\alpha})
\end{equation}
 changes the gauge fixed fermion only and can be the symmetry of the generating functional for 1-form theory in BV formulation
\begin{equation}
Z[\Phi^*]\stackrel{\stackrel{FFBRST with}{------------\longrightarrow}}{\Theta ^\prime}Z[\tilde{\Phi}^*]
\end{equation}
In 2-form gauge theory we can construct the finite field anti-BRST transformations parameter corresponding to
\begin{eqnarray}
\Theta^\prime_{ab} &=&-\int d^4x\left[\gamma_1\rho_\nu (\partial_\mu B^{\mu\nu}-\eta_\mu B^{
\mu\nu}-\partial^\nu\varphi -\eta^\nu\varphi )+\gamma_2 \lambda_1\rho_\nu\beta^\nu \right. 
\nonumber\\
&-&\left.\gamma_1\sigma (\partial_\mu\tilde\rho^\mu -\eta_\mu\tilde\rho^\mu )+\gamma_2 
\lambda_2\sigma\tilde\chi\right],
\end{eqnarray}
is the symmetry generating functionals in BV formulation of 2-form gauge theory.  Thus, FFanti-BRST transformations with appropriate parameters are also the symmetry of the generating functionals BV        formulation.

\begin{center}
{\bf {\Large {Conclusion}}}
\end{center}
Generalized BRST transformations in which the parameter is finite and field dependent are also the symmetry of FP effective action and are nilpotent. In path integral formulation of different gauge theories, such FFBRST transformations do not leave the path integral measure in the definition of generating functionals invariant and hence the generating functionals are not invariant under such transformations. In fact, with appropriate choice of the finite field dependent parameter, these transformations are shown to relate different generating functionals corresponding to different effective gauge theories. Because of this important results, FFBRST transformations found many applications in the study of gauge field theories. In the present work, we showed that such transformations leaves the generating functionals defined in BV formulation of different gauge theories invariant. We chose the finite field dependent parameter in such a manner that the contribution from non-trivial Jacobian of path integral measure is absorbed in the expression of gauge fixed fermions in BV formulation. Recalling the well known result in BV formulation that the generating functionals are independent of the choice of gauge fixed fermion, we claim that FFBRST transformations with appropriate parameter are the formal symmetry of the generating functionals defined in BV formulation. In principle, one can construct infinitely many such parameter for which FFBRST leaves the generating functionals invariant in BV formulation. We have constructed parameters both for 1-form and 2-form gauge theories  in this work. However, the disadvantage of such transformations is that these transformations  are non-local. We believe that such transformations will also leave the quantum master equation invariant in BV formulation which further can lead to important consequences.

{\Large{\bf {Acknowledgment}}} \\

\noindent
We thankfully acknowledge the financial support
from the Department of Science and Technology (DST), Government of India, under
the SERC project sanction grant No. SR/S2/HEP-29/2007.\\

\end{document}